\documentclass{elsart}

\usepackage{graphicx}

\newcommand{\beq}{\begin{equation}}
\newcommand{\eeq}{\end{equation}}
\newcommand{\eq}[1]{Eq.~(\ref{#1})}
\newcommand{\nueq}[1]{Eqs.~(\ref{#1})}
\newcommand{\nnueq}[1]{(\ref{#1})}

\begin{document}

\begin{frontmatter}

\title{The g factor of proton}

\author[vniim,mpq]{Savely G. Karshenboim}
\ead{sek@mpq.mpg.de}
\and
\author[gao,vniim]{Vladimir G. Ivanov}
\address[vniim]{D. I. Mendeleev Institute for Metrology (VNIIM),
 St. Petersburg 198005, Russia}
\address[mpq]{Max-Planck-Institut f\"ur Quantenoptik, 85748 Garching, Germany}
\address[gao]{Pulkovo Observatory, 196140, St. Petersburg, Russia}

\begin{abstract}
We consider higher order corrections to the $g$ factor of a bound proton
in hydrogen atom and their consequences for a magnetic moment of free
and bound proton and deuteron as well as some other objects.
\end{abstract}


\end{frontmatter}

Investigation of electromagnetic properties of particles and nuclei 
provides important information on fundamental constants. In addition,
one can also learn about interactions of bound
particles within atoms and interactions of atomic (molecular) composites
with the media where the atom (molecule) is located. Since the magnetic
interaction is weak, it can be used as a probe to learn about atomic and
molecular composites without destroying the atom or molecule. In
particular, an important quantity to study is a magnetic moment for 
either a bare nucleus or a nucleus surrounded by electrons.

The Hamiltonian for the interaction of a magnetic moment {\boldmath$\mu$} 
with a homogeneous magnetic field ${\bf B}$ has a well known form
\beq
H_{\rm magn} = - {\mbox{\boldmath$\mu$}}\cdot{\bf B}\;,
\eeq
which corresponds to a spin precession frequency
\beq\label{omega}
  h\nu_{\rm spin} = \frac{\mu}{I} B
  \,,
\eeq
where $I$ is the related spin equal to either 1/2 or 1 for
particles and nuclei under consideration in this paper. Comparison of
the frequencies related to two different objects allows to exclude the
magnetic field $B$ from equations and to determine the ratio of their
magnetic moments with an accuracy sometimes substantially higher than 
that in the determination of the applied magnetic field.

To measure the magnetic moment of a given nucleus one has to
compare it with a value of some probe magnetic moment, which should be
known or determined separately. For a significant number of most accurate
measurements the probe value is related to the magnetic moment of a free
or bound proton and a crucial experiment on its determination is related
to a proton bound in hydrogen atom \cite{codata,wink}. 
Nuclear magnetic moments are usually presented in units of nuclear magneton ($\mu_N$)
\cite{adndt,fire}, which is related to the proton magnetic moment via the
relation
\beq
  \mu_p = \frac{1}{2} g_p \mu_N
  \,,
\eeq
where $g_p$ is the proton $g$ factor and $\mu_N=e\hbar/2m_p$.

The spin precession frequency was
studied not only for a free proton, but also for the one bound in atoms or
molecules located in gaseous or liquid media. The magnetic moment and the
$g$ factor of a bound proton differ from their free values (see e.g.
\cite{breit,BS}). The purpose of this paper is to re-evaluate in part available 
experimental data for light atoms and in particular to determine 
the $g$ factor of a free proton ($g_p$) and a proton bound in the 
ground state of hydrogen atom ($g_p({\rm H})$) from experiment \cite{wink}. We
also study the consequences of re-evaluaton of $g_p$ and similar experiments for 
deuterium \cite{phil84} and muonium \cite{liu99}.

\begin{figure}
\begin{center}
\includegraphics[width=0.6\textwidth]{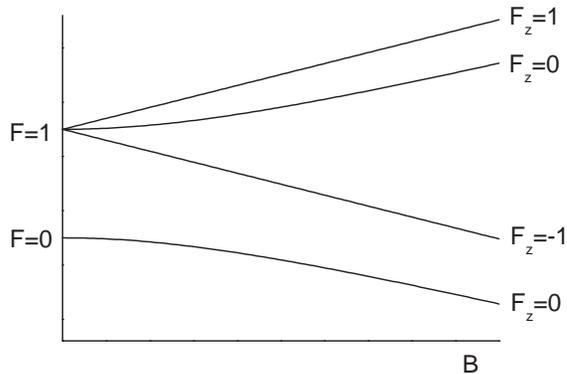}
\end{center}
\caption{Sublevels $E_{\rm magn}(F,F_z)$ of the hyperfine
structure in the ground state of hydrogen atom in a homogenous 
magnetic field $B$.} \label{mfFig}
\end{figure}

The most accurate determination of the $g$ factor of a free proton was
performed studying the hyperfine structure of the hydrogen atom in the
homogeneous magnetic field. The dependence of hyperfine sublevels of the
ground state in the hydrogen atom on the value of the magnetic field  
${\bf B}$ directed along the $z$ axis is shown in Fig.~\ref{mfFig} (see e.g.
\cite{BS}). The energies of hyperfine components $E_{\rm magn}(F,F_z)$ of
the $1s$ state are described by
\beq\label{levels}
  \begin{array}{l@{\,=\,}l}
  E_{\rm magn}(1,+1) & \frac{1}{2}\bigl(E_e - E_p\bigr) + \frac{1}{4}E_{\rm hfs} \,,\\
  E_{\rm magn}(1,0) & \frac{1}{2} \sqrt{(E_e+E_p)^2+E_{\rm hfs}^2}
    - \frac{1}{4}E_{\rm hfs} \,,\\
  E_{\rm magn}(1,-1) & - \frac{1}{2}\bigl(E_e - E_p\bigr) + \frac{1}{4}E_{\rm hfs} \,,\\
  E_{\rm magn}(0,0) &  -\frac{1}{2}\sqrt{(E_e+E_p)^2+E_{\rm hfs}^2}
    - \frac{1}{4}E_{\rm hfs} \,,\\
  \end{array}
\eeq
where $E_e=g_e(H)\mu_BB$ and $E_p=g_p(H)\mu_NB$ are related to precession
frequencies of electron and proton, and $\mu_B=e\hbar/2m_e$ is the Bohr
magneton. The energy splitting $E_{\rm hfs}$ is related to the hyperfine
interval in the hydrogen ground state known with high accuracy in frequency
units.

The experiment \cite{wink} devoted to a measurement of splitting and shift
of the hyperfine sublevels in hydrogen atom due to the magnetic field
led to the following result \cite{codata}\footnote{Here and further we ignore the direction of spin and magnetic moment and thus the sign of some $g$ factors and ratios of magnetic moments.}
\beq\label{hdatum}
  \frac{\mu_e({\rm H})}{\mu_p({\rm H})}
  = \frac{g_e({\rm H})}{g_p({\rm H})}\;\frac{\mu_B}{\mu_N}=
  658.210\,705\,8(66)
  \,.
\eeq

The result for the related ratio of the free magnetic moments derived 
in the experiment \cite{wink} and quoted in Ref.~\cite{codata} was based 
on a theoretical expression which contained relativistic and recoil 
corrections up to the third order in either of three parameters, such 
as the free QED parameter $\alpha$ (appearing due to the anomalous 
magnetic moment of electron), the strength of the Coulomb potential 
$Z\alpha$ and the recoil parameter $m_e/m_p$. All these terms are of 
pure kinematic origin and
were derived before (see, e.g., Ref~\cite{th}). The $g$ factors in 
hydrogen atom in terms of the free $g$ factors of the electron
\cite{vand}
\beq\label{geco}
g_e = 2.002\,319\,304\,376(8)
\eeq
and the proton ($g_p$) are
\begin{eqnarray}
  \label{ge}
  g_e({\rm H}) &=& g_e  \cdot\left\{1-\frac{(Z\alpha)^2}{3}
 \left[   1-\frac{3}{2}\frac{m_e}{m_p} \right]
 +\frac{\alpha(Z\alpha)^2}{4\pi} \right\}  \,,\\
  \label{gp}
  g_p({\rm H}) &=&  g_p\cdot
  \left\{1- \frac{\alpha(Z\alpha)}{3}\left[ 1 - \frac{m_e}{2m_p}
  \frac{3+4a_p}{1+a_p} \right] \right\}\,,
\end{eqnarray}
where for the anomalous magnetic moment of the proton we set $a_p\simeq 1.792\,847...$. This value is taken from Ref.~\cite{codata}. It enters only small corrections (see e.g. \eq{gp}) and any re-evaluation which can shift the proton $g$ factor on level of a part in $10^8$ will not affect those corrections.

These expressions applied to evaluations in 
Refs.~\cite{codata,wink} include only the terms up to the third order.
However, for the determination of the ratio of the magnetic moments at
the level of a part in $10^8$ the higher order corrections should be 
taken into account as well.
The fourth order corrections are in part nuclear-spin-dependent. E.g., in
the case of hydrogen atom ($I=1/2$) the expression for higher-order
terms corrections reads (cf. Ref.~\cite{ki_cjp})
\begin{eqnarray}
  \Delta g_e({\rm H}) &=&
  g_e \cdot
  \left\{
   - \frac{(Z\alpha)^2(1+Z)}{2}\left(\frac{m_e}{m_p}\right)^2
   - \frac{5\alpha(Z\alpha)^2}{12\pi}\frac{m_e}{m_p}
 \right.\nonumber\\&&\phantom{g_e\cdot\biggl\{}\left.
    -\bigl(0.289\dots\bigr)\times\frac{\alpha^2(Z\alpha)^2}{\pi^2}
 -\frac{(Z\alpha)^4}{12}
\right\}
 \,,\label{geh}\\
  \Delta g_p({\rm H})&=&
  g_p
  \cdot\left\{
    \alpha(Z\alpha)\left(\frac{m_e}{m_p}\right)^2
    \left(
      -\frac{1}{2} - \frac{Z}{6}\frac{3-4a_p}{1+a_p}
    \right)
    -\frac{97}{108}\,\alpha(Z \alpha)^3
  \right\}
  \,.\label{gph}
  \end{eqnarray}

After a proper substitution for $m_p$, $g_p$ and $a_p$, the results 
for the leading terms in \nueq{ge} and \nnueq{gp} can be applied to any hydrogen-like atoms, while
the higher-order corrections in \nueq{geh} and \nnueq{gph} can be used only in the case of the nuclear spin 1/2 (e.g. for the tritium atom and a hydrogen-like helium-3 ion). 
For the deuterium atom ($I=1$) the results 
for the higher-order terms differ from \nueq{ge} and \nnueq{gp} 
and have to be 
properly corrected. E.g., following Ref.~\cite{marty}, we obtain
\begin{eqnarray}
  \Delta g_e({\rm D}) &=&
  g_e \cdot
  \left\{
   - \frac{(Z\alpha)^2(11Z+12)}{18}\left(\frac{m_e}{m_d}\right)^2
   - \frac{5\alpha(Z\alpha)^2}{12\pi}\frac{m_e}{m_d}
 \right.\nonumber\\&&\phantom{g_e\cdot\biggl\{}\left.
  -\bigl(0.289\dots\bigr)\times\frac{\alpha^2(Z\alpha)^2}{\pi^2}
  -\frac{(Z\alpha)^4}{12}
\right\}
 \,,\label{ged}\\
  \Delta g_d({\rm D})&=&
  g_d
  \cdot\left\{
    \alpha(Z\alpha)\left(\frac{m_e}{m_d}\right)^2
    \left(
     - \frac{1}{2} -\frac{Z}{3}\frac{2-2a_d}{1+a_d} 
    \right)
    -\frac{97}{108}\,\alpha(Z \alpha)^3
  \right\}
  \,,\label{gpd}
\end{eqnarray}
where 
\begin{equation}
\mu_d = g_d\mu_N 
\equiv (1+a_d)\frac{e\hbar}{m_d}
\end{equation}
and $a_d\simeq - 0.142\,987...$

There is only a single experiment \cite{geged} where the recoil part 
of the higher-order terms in \nueq{geh} and \nnueq{ged} is important. 
In this experiment the electron magnetic moments of hydrogen and 
deuterium \cite{geged} were compared. In contrast, the non-recoil 
higher-order terms in \nueq{geh} and \nnueq{ged} are not important 
for this isotopic comparison.
The accuracy of a similar experiment on hydrogen and tritium \cite{geget} 
was not high enough to be sensitive to the recoil corrections in \eq{geh} .

An opposite situation appears in the experimental comparison of the 
nuclear magnetic moment and the electron magnetic moment while studying 
e.g. the hydrogen energy levels in
\eq{levels}. We note that only one higher-order correction for each $g$
factor can contribute at a level close to a part in $10^8$
\begin{eqnarray}
  \Delta g_e( {\rm H} ) &=&
  -\frac{(Z\alpha)^4}{12} \cdot g_e
   \,,\label{geh1}\\
  \Delta g_p({\rm H})&=&
   -\frac{97}{108}\,\alpha(Z \alpha)^3
\cdot g_p
   \,.\label{gph1}
\end{eqnarray}
The former equation owing a small numerical coefficient $1/12$ is related to a smaller effect ($\Delta g_e/g_e \simeq  -2.4\times10^{-10}$). It has been known 
for a while \cite{breit} and was an only fourth-order term included into evaluation in Ref.~\cite{codata}, while the latter correction ($\Delta g_p/g_p \simeq  -2.6\times10^{-9}$) was
obtained recently \cite{moor,pyp,ki_cjp}. Thus, the higher-order recoil
effects can be neglected and that is fortunate because the remaining 
terms in \nueq{geh1} and \nnueq{gph1} are nuclear-spin-independent.

Combining \nueq{geco}, \nnueq{ge} and \nnueq{geh} we find for the hydrogen atom
\beq \label{muehB}
\frac{1}{2} g_e({\rm H} ) =  \frac{\mu_e({\rm H} )}{\mu_B}
  = 1.001\,141\,926\,3\,,
\eeq
where the uncertainty is below a part in $10^{10}$ and can be neglected
in further considerations.

Applying the results for the higher-order corrections from \nueq{gph1} 
and \nnueq{muehB} to the experimental data in \eq{hdatum} \cite{wink,codata}, 
we deduce
\beq\label{muhn}
  \frac{\mu_p({\rm H})}{\mu_B} = 0.001\,521\,005\,230(15)\;,
\eeq
\beq
  \frac{\mu_p}{\mu_B} = 0.001\,521\,032\,207(15)
\eeq
and
\beq
  \frac{\mu_p}{\mu_e} = 658.210\,685\,9(66)\,.
\eeq

To interpret the results in units of the nuclear magneton, we have to apply
an accurate value of the conversion factor
\beq
  \frac{\mu_B}{\mu_N} = \frac{m_p}{m_e}
  \,.
\eeq
The proton-to-electron mass ratio was recently determined from an experiment
on the $g$ factor of a bound electron in hydrogen-like carbon \cite{haef}
and the result \cite{beier,km_plb} is more accurate, being slightly
different from the one based on comparison of cyclotron frequencies of
electron and proton \cite{farn,codata}. We note that this new approach to
the determination of the electron-to-proton mass ratio \cite{h2,beier} was
confirmed by a measurement of the $g$ factor of a bound electron in the
hydrogen-like oxygen \cite{werth_cjp}, as suggested in Ref.~\cite{kis}.
The experimental result \cite{werth_cjp} is in fair agreement with theory
\cite{kis,ki_cjp,km_plb}. Other less accurate results on the
proton-to-electron mass ratio are overviewed in Ref.~\cite{ki_cjp}.

The values of the electron-to-proton mass ratio deduced from experiment \cite{haef} are slightly different from evaluation to evaluation, and here we use the one found in Ref.~\cite{ki_cjp} (see also discussion in Ref.~\cite{km_plb})
\beq\label{factor}
  \frac{\mu_B}{\mu_N} = \frac{m_p}{m_e} = 1\,836.152\,673\,6(13)
  \,.
\eeq
Using a value for the magnetic moment of a bound proton from \eq{muhn} we
arrive to the following results for the proton $g$ factor
\beq
  \frac{g_p({\rm H})}{2} = \frac{\mu_p({\rm H} )}{\mu_N} = 2.792\,797\,820(28)
\eeq
and
\beq
  \frac{g_p}{2} = \frac{\mu_p}{\mu_N} = 2.792\,847\,353(28) \,.
\eeq

For further application we also need a value of the electron magnetic
moment in hydrogen atom in units of the nuclear magneton. Combining
\nueq{muehB} and \nnueq{factor}, we obtain
\beq \label{mueh}
  \mu_e({\rm H} )
    =  1838.249\,424\,6(13)\, \mu_N
  \,.
\eeq

Similar analysis can be performed for experiments with the deuterium atom.
The experimental result for deuterium \cite{phil84} reads
\beq
  \frac{\mu_e({\rm D} )}{\mu_d({\rm D} )}= 2143.923\,565(23)\;.
\eeq
Taking into account higher-order corrections in Eqs. (\ref{geh1}) and
(\ref{gph1}), we obtain
\beq
  g_d({\rm D}) = \frac{\mu_d({\rm D})}{\mu_N} = 0.857\,423\,017\,1(94)
\eeq
and
\beq
  g_d = \frac{\mu_d}{\mu_N} =  0.857\,438\,233\,3(94)\,.
\eeq

Let us consider some consequences of correcting the $g$ factor of a
free proton and magnetic moments of a proton and electron bound in the hydrogen atom. E.g.,
the $g$ factor of a shielded proton in water was measured \cite{phil77} in
comparison with the magnetic moment of an electron bound in hydrogen
atom (\ref{mueh}). The corrected results for the proton magnetic moment
are
\beq
  \frac{\mu^\prime_p(\mbox{H$_2$O})}{\mu_e} = 0.001\,520\,993\,127(17)
\eeq
and
\beq\label{gph2o}
  \frac{g_p^\prime(\mbox{H$_2$O})}{2}  = \frac{\mu_p^\prime(\mbox{H$_2$O})}{\mu_N} =
  2.792\,775\,600(33) \,,
\eeq
where shielding is denoted by prime. These results are related to a spherical
sample of pure water at a temperature $t=25^\circ {\rm C}$. The values for
other forms and temperatures of the sample can be recalculated (for
detail see Ref.~\cite{codata} and references therein).

The corrections for a bound proton in water is shifted by approximately 30\% from the original result, however, the difference between the results from Ref.~\cite{codata} and ours has been reduced since the former evaluation included a result from \cite{gamma} which is ten times less accurate and about two standard deviations off from the more accurate value \cite{phil77}. Here, we consider only most accurate results while the other data and in particular the result \cite{gamma} have been dismissed from our consideration.

A determination of the magnetic moment of the proton in water is
important because it has been used as a probe in a number of measurements
and in particular to determine a value of the magnetic moment of a
shielded helion, a nucleus of the $^3$He atom, \cite{flow}
\beq
  \frac{\mu^\prime_{h}({}^3{\rm He})}{\mu^\prime_p({\rm H}_2{\rm O})}
  = 0.761\,786\,131\,3(33)\,.
\eeq
With a corrected value for the magnetic moment of the shielded proton in
\eq{gph2o} the helion result now reads
\beq
  \frac{\mu^\prime_h(^3{\rm He})}{\mu_N} = 2.127\,497\,720(25)\,.
\eeq

The magnetic moment $\mu^\prime_h(^3{\rm He})$ is related to a helion
bound in a neutral atom and studied in a low pressure helium-3 gas. To the
best of our knowledge, no theoretical calculations are available for
the higher-order correction to the $g$ factor $g^\prime_h(^3{\rm He})$
similar to the $\alpha(Z\alpha)^3$ in Eq.~(\ref{gph1}) for hydrogen.
The single-electron contribution for helium should be doubled (because of the presence of two
electrons) and could receive some enhancement since the
effective charge for each electron is somewhat bigger than unity. There
should also be some essentially two-electron relativistic effects.
We expect that the uncertainty of any theoretical calculation (see
e.g. \cite{neronov}), ignoring the higher-order relativistic effects in
order $\alpha(Z\alpha)^3$, cannot be below a part or even a few parts in
$10^8$. Because of the unclear status of the uncertainty of theoretical
calculations of the screening effects, we do not consider here a
determination of the free nuclear magnetic moment of helion.

We have also considered data related to the muon magnetic
moment. The result 
\beq
\frac{\mu_\mu}{\mu_N}= 8.890\,596\,96(42)
\eeq
is a weighted average of two values:
\begin{itemize}
\item
The first one (${\mu_\mu}/{\mu_N} = 8.890\,597\,05(106)$) is obtained from
the measurement \cite{liu99} of the transitions between hyperfine components
of the ground state in muonium (cf. Fig.~\ref{mfFig}) in the magnetic
field calibrated by measuring precession of a free proton. The value
was slightly corrected in \cite{ki_cjp} because of higher-order
corrections\footnote{We note that the $\alpha^2(Z\alpha)m/M$ term in Eq.
(9) of Ref. \cite{ki_cjp} is to be corrected and it now reads
$ \alpha^2(Z\alpha)/12\pi\;m/M$.}.
\item
The other result (${\mu_\mu}/{\mu_N} = 8.890\,596\,95(46)$) is found from a
value of the hyperfine splitting in the muonium ground state measured for
zero magnetic field \cite{liu99} and compared with theory \cite{cek}.
Note that the fine structure constant used for the calculations here is
$\alpha^{-1} = 137.035\,998\,76(52)$ \cite{kino}.
\end{itemize}
The less accurate data on the muon magnetic moment have  been overviewed
in Ref.~\cite{ki_cjp} in terms of a related quantity
\beq
\frac{m_e}{m_\mu} = \frac{1}{1+a_\mu}\frac{\mu_\mu}{\mu_N}
\frac{\mu_N}{\mu_B}\;.
\eeq
They are statistically not important and have not been taken 
into account while calculating the muon result above.

\begin{table}
\caption{Magnetic moments ratios of electron, muon, proton, deuteron and
helion. The CODATA results are taken from the {\em CODATA Recommended
values -- 1998} \protect\cite{codata} and the {\em corrected} results are
discussed in our paper. We restore here signs of magnetic moments.\label{Tab}}
\begin{center}
\begin{tabular}{lll}
\hline\hline
\multicolumn{1}{c}{Value} & \multicolumn{1}{c}{CODATA
\protect\cite{codata}} & \multicolumn{1}{c}{Corrected}\\
\hline
$\mu_B/\mu_N$ & 1836.152\,667\,5(39) & 1836.152\,673\,6(13)\\
$\mu_e/\mu_N$ & -- 1838.281\,966\,0(39)& -- 1838.281\,972\,1(13)\\
$\mu_e({\rm H})/\mu_N$ & -- 1838.249\,418\,7(39) & -- 1838.249\,424\,6(13)\\
$\mu_\mu/\mu_N$ & -- 8.890\,597\,70(27) & -- 8.890\,596\,96(42)\\
\hline
$\mu_p/\mu_N$ & 2.792\,847\,337(29)& 2.792\,847\,353(28) \\
$\mu_p({\rm H})/\mu_N$ & 2.792\,797\,812(29) & 2.792\,797\,820(28)\\
$\mu^\prime_p(\mbox{H$_2$O})/\mu_N$ & 2.792\,775\,597(31) & 2.792\,775\,600(33)\\
$\mu_d/\mu_N$ & 0.857\,438\,228\,4(94) & 0.857\,438\,233\,3(94)\\
$\mu_d({\rm D})/\mu_N$ & 0.857\,423\,014\,4(94) & 0.857\,423\,017\,1(94)\\
$\mu^\prime_h(^3{\rm He})/\mu_N$ & -- 2.127\,497\,718(25) & -- 2.127\,497\,720(25)\\
\hline\hline
\end{tabular}
\end{center}
\end{table}

To summarize our consideration, we present the corrected values of the $g$
factors of electron (bound), proton (free and bound), deuteron (free and
bound) and helion (bound) in Table~\ref{Tab}. We compare our results with those in Ref.~\cite{codata} which seems to be the only paper where a systematic consideration on theory and experiment in light simple atoms is done.
We resume that higher-order corrections are somewhat bigger than it was expected in  Ref.~\cite{codata} but still do not exceed the uncertainty.
In particular, the corrections to the proton, deuteron and helion magnetic moments ($g$ factors) are slightly below the uncertainty, which is for all these quanitites on level of a part in $10^8$ in fractional units. However the corrections are important because they produce a systematic effect on deduced values of all discussed nuclear magnetic moments at a level of an essential part of uncertainty. A shift of the value of the nuclear magnetic moments listed in Table~1 typically varies from 30\% to 60\% of the value of their uncertainties. Note, that in the case of the magnetic moment of the proton in water and a related value of the helion magnetic moment the shift is still on a level of 30\% of the uncertainty, but it corresponds to a result of the most accurate experiment \cite{phil77}, while the CODATA result in Table~1 is related to an averadge value (see discussions after \eq{gph2o}). 

The authors would like to thank Gordon Drake, Simon Eidelman, Jim Friar, Peter Mohr  
and Yurii Neronov for useful and stimulating discussions. The work was supported in part by RFBR under grants 00-02-16718, 02-02-07027 and 03-02-16843.

\end{document}